\documentclass{ws-procs10x7}
\usepackage{balance}
\usepackage{graphicx}
\usepackage{amsmath}
\newcolumntype{d}[1]{D{.}{.}{#1}}

\def\Journal#1#2#3#4{{\it #1} {\bf #2}, #3 (#4)}

\makeindex
\begin{document}

\title{Quantum Nernst Effect in a Bismuth Single Crystal}

\author{MARI MATSUO$^*$ }
\address{Graduate School of Integrated Sciences, Ochanomizu University, 
Otsuka, Bunkyo, Tokyo 112-8610,  Japan\\$^*$E-mail:  matsuo@iis.u-tokyo.ac.jp}

\author{AKIRA ENDO}
\address{Institute for Solid State Physics, University of Tokyo, Kashiwanoha, Kashiwa, Chiba 277-8581,  Japan}

\author{NAOMICHI HATANO}
\address{Institute of Industrial Science, University of Tokyo, Komaba, Meguro, Tokyo 153-8505, Japan}

\author{HIROAKI NAKAMURA}
\address{Department of Simulation Science, National Institute for Fusion Science, Oroshi-cho, Toki, Gifu  509-5292, Japan}

\author{RY$\overline{{\rm O}}$EN SHIRASAKI}
\address{Department of Physics, Faculty of Engineering, Yokohama National University,
Tokiwadai, Hodogaya-ku, Yokohama, Kanagawa 240-8501, Japan}

\author{KO SUGIHARA}
\address{1-40-6-506 Shibayama, Funabashi, Chiba, 274-0816, Japan}

%%%%%%%%%%%%%%%%%%%%%%%%%%%%%%%%%%%%%%%%%%%%%%%%%%%%%%%%%%%%%%%%%%%%%%%%%
% You may repeat \author \address as often as necessary                 %
%%%%%%%%%%%%%%%%%%%%%%%%%%%%%%%%%%%%%%%%%%%%%%%%%%%%%%%%%%%%%%%%%%%%%%%%%
%\renewcommand{\baselinestretch}{2}
\twocolumn[
\maketitle\abstract{
We report a theoretical calculation explaining the quantum Nernst effect observed experimentally in a bismuth single crystal. 
Generalizing the edge-current picture in two dimensions, we show that the peaks of the Nernst coefficient survive in three dimensions due to a van Hove singularity. 
We also evaluate the phonon-drag effect on the Nernst coefficient numerically. 
Our result agrees with the experimental result for a bismuth single crystal.
}
]
%%%%%ここまでアブストラクト
%%%%%ここから本文
%%%%%%イントロ%%%%%%%%%
\section{Introduction}
The Nernst effect is a thermoelectric phenomenon which yields a transverse voltage when there is a magnetic field perpendicular to a temperature gradient.
In 2005, Nakamura {\it et al}.\cite{nernst} predicted the {\it quantum} Nernst effect of the two-dimensional electron gas in a semiconductor heterojunction under a strong magnetic field, using the edge current picture of the quantum Hall effect.\cite{Halperin}
In the quantum Nernst effect, 
the Nernst coefficient shows sharp peaks when a Landau level crosses the Fermi level  
 and the thermal conductance in the direction of the temperature gradient is quantized.
Shirasaki {\it et al.}\cite{shirasaki} showed that impurities do not change the conclusions much. 

In 2007, Behnia {\it et al.}\cite{Behnia,BehniaScience} have reported quantum oscillation of the Nernst coefficient in a bismuth single crystal, which was similar to the prediction by Nakamura {\it et al}.\cite{nernst} 
It is remarkable that the strong quantum effect is observable in three dimensions; 
the quantized plateaus of the Hall conductance can be hardly identified in three dimensions.

 %%本研究は、、、、
In the present study we extend the theoretical prediction of the quantum Nernst effect in two dimensions\cite{nernst} to a three-dimensional system, 
motivated by the experiment in a bismuth single crystal.
We find that the peaks of the Nernst coefficient survive in three dimensions because of a van Hove singularity. 
A simple calculation explains Behnia {\it et al.}'s experimental result qualitatively.
Consideration of the electron-phonon interaction leads even to quantitative agreement with the  experimental result.
%The results have been compared with those obtained with 
%van Hove singularity***** 

%以前の研究のレビュー
%Thermoelectric transport in bismuth have studied both experimentally\cite{Farag}, and theoretically\cite{Koren}.
%Kubo et at. developed　the linear transport theory for strong magnetic field condition \cite{Kubo}.  

%表現
%In recent years,
%Additionally, we determined that it is possible that 
%Here, we further investigate these ~ by extending our calculations to ~

%ビスマスのこと
%Bismuth crystal has a rhombohedral structure and has both electron and hole as carrier; semimetal.
%We considered that the adiabatic Nernst effect for a magnetic field along trigonal (we set z axis) at low temperatures, low enough for mean free path to be greater than the system size.
%For simplicity, we assume that the shape of  Fermi surface of bismuth is ellipsoid, there is a only hole as carrier in bismuth.

%effective mass   
%楕円型のフェルミ面m、mz
%z方向に磁場をかける　トリゴナル
% キャリアはホールのみ
% 不純物なし＝平均自由行程が長い系の長さより
% for a field along the trigonal axis at low temperatures.

%%%%%端電流%%%%%%%%%%%%%%%%%%%%
\section{Edge-current picture of the quantum Nernst effect}
%断熱ネルンスト
We consider the adiabatic Nernst effect. 
%in a clean sample at low temperatures, clean and low enough for the mean free path to be greater than the system size. 
The magnetic field $B$ is applied in the direction of the trigonal axis, which we set to the $z$ axis.
The system is electrically and thermally insulated on all surfaces except for the two surfaces attached to heat baths, which induce a temparature bias $\Delta T$ in the $x$ direction.
Then the Nernst voltage $V_N$ is generated in the $y$ direction.
The adiabatic Nernst coefficient is then defined by
\begin{equation}
N = - \frac{V_N/ W}{B \Delta T / L}, 
\end{equation}
where $L$ and $W$ are the lengths of the crystal in the $x$ and $y$ directions, respectively.

%%考える系
%Our model 
%bismuth single crystal
%at low temperature
%applied strong magnetic field along to z-axis
%↑↑↑もう言ったので必要なし

%３次元の系の絵はもういらない
%\begin{figure}[htb]
%%\begin{center}
%\includegraphics[width=2.2in]{nernst3D.eps}
%%\end{center}
%\caption{A schematic view of}
%\label{graph}
%\end{figure}
% 

%理論と実験を比較してみる
%attempt a comparison between theory and experiments

%断熱ネルンスト
%端電流描像
%フォノンドラッグ
%仮定：フェルミ面は楕円体（mとmz）／キャリアはホール１種類としました
%不純物なし／低温強磁場中／

%線形応答の範囲でみている
%We construct the linear transport equations of the thermomagnetic transport using~, which we obtain within ~approximation.
%In the regime of linear response, this is conventionally expressed as
Nakamura {\it et al.}\cite{nernst} assumed that, in a two-dimensional clean sample at low temperatures, an edge current circulates around the system ballistically when the chemical potential is in between two Landau levels.
The edge current that departs the colder side maintains the Fermi distribution $f(T_-,\mu_-)$ of the colder bath 
until it reaches the hotter side 
and then maintains the Fermi distribution $f(T_+,\mu_+)$ of the hotter bath 
until it reaches the colder side.
The difference of the chemical potential generates the Nernst voltage as $V_N = (\mu_+ - \mu_-) / |e|$.

The Hamiltonian of the two-dimensional electron gas yields energy levels $E(n,k_x)$, where $n = 0,1,2, \cdots$ represents a Landau level and the label $- k_{\rm m} \le k_x \le k_{\rm m}$ represents a state in the Landau level with $k_{\rm m}$ determined by the confining potential in the $y$ direction.
After some algebra they obtained the electric current in the $x$ direction as
\begin{equation}
I = \frac{e}{\pi \hbar}
\left(
 A_{0} \Delta \mu +  A_{1}  k_B \Delta T
\right) , 
\end{equation}
where $\Delta T \equiv T_+ - T_- , \,  \Delta \mu \equiv \mu_+ - \mu_-$ and
\begin{equation}
A_{\nu}(\mu) = \sum_{n=0}^{\infty}  \int_{X_0(n)}^{X_1(n)} \frac{x^{\nu} dx}{4 \cosh^2 (x/2)}
\label{A}
\end{equation}
with $X_i(n) \equiv (E_i(n) - \mu)/(k_BT), \, 
E_0(n) \equiv E(n,k_x = 0)$ and $E_1(n) \equiv E(n,k_x = k_{\rm m})$.
Setting $I = 0$ and $e = |e|>0$, they obtained the Nernst coefficient as\cite{nernst}
\begin{equation}
N(B,T) = \frac{k_B}{|e| B_z} \frac{L}{W} \frac{A_1}{A_0},\label{nernstA}
\end{equation}
which showed sharp peaks when the chemical potential is equal to a Landau level. 
This quantum behavior basically comes from the oscillation of the coefficient $A_1$.

%熱伝導度の式
%and,  similarly we obtain the thermal conductance
%\begin{equation}
%G_Q \equiv \frac{I_Q}{\Delta T} = \frac{k_B^2 T}{\pi \hbar} 
%\left[ A_2 - \frac{(A_1)^2}{A_0}
%\right]
%\end{equation}

%%低温、バリスティック、線形、ΔT、Δμ小さい、平均自由行程長い、不純物なし　を入れること

%%計算してみました
%We evaluated the Nernst coefficient N and the thermal conductance $G_Q$ as in Fig.~\ref{nernst}.
%using a confinement potential $V(y)$ as follows.

%%こういうポテンシャルです($w$や$V$の値)
%\begin{eqnarray}
%V(y) = 
%\begin{cases}
%0 & (|y| \le  \frac{w}{2}) \\
%\frac{m \omega_0^2}{2} \left( |y| - \frac{w}{2} \right)  &  
% \left( \frac{w}{2} <  |y| < \frac{W}{2 }\right)
%\end{cases}
%\end{eqnarray}

%ここから今回の３次元
We now extend the above to three dimensions.
This extension introduces an additional argument $k_z$ in the energy level as
$E_i(n,k_z) = E_i(n) + (\hbar k_z)^2/(2 m_z)$
and hence in $X_i(n,k_z)$.
Equation (\ref{A}) now has integration over $k_z$ as
\begin{eqnarray}
\tilde{A_{\nu}}(\mu) 
&=& \sum_{n=0}^{\infty} \int \frac{d k_z}{2\pi}
 \int_{X_0(n,k_z)}^{X_1(n,k_z)} \frac{x^{\nu} dx}{4 \cosh^2 (x/2)} \nonumber \\
% &=&\sum_{n=0}^{\infty} \int d E_z D(E_z) 
% \int_{X_0(n) + E_z/(k_B T)}^{X_1(n)+ E_z/(k_B T)} \frac{x^{\nu} dx}{4 \cosh^2 (x/2)} 
% \nonumber \\
 &=& \int_0^{\infty} D(E_z) A_{\nu} (\mu - E_z) dE_z ,
\end{eqnarray}
where we have changed the integration variable from $k_z$ to $E_z = (\hbar k_z)^2/(2m_z)$. 
This variable transformation introduces the density of states 
$D(E_z) \equiv \left[ 2 \pi (d E_z / d k_z) \right]^{-1}$, 
which has a van Hove singularity at $E_z=0$.

This van Hove singularity is the essence of the survival of the quantum oscillation in three dimensions. 
The overlap of the peaks of $A_1(\mu)$ with the van Hove singularity leads to oscillatory behavior of the coefficient $\tilde{A_1}(\mu)$ and hence of the Nernst coefficient.

%%%van Hove singularityの議論を書きましょう♪
%Since in two dimensions the density of state represents van Hove singularities because  $(d E_{xy}(n,k_x)/ d k)$ is vanish
%when the chemical potential coincide with one of Landau levels,
%the thermal conductance shows plateaus and the Nernst coefficient shows the peaks.
%Extending this idea to three dimensions, 
%the van Hove singularity is superimposed,
%since  there is the density of state  ,because of $E_z = (\hbar k_z)^2 / (2 m_z)$,
%even at  the place which it shifts from the place 
% where the chemical potential  is corresponding to one of Landau levels in x-y plain
%Consequently
%the peaks of the Nernst coefficients survive, which shows quantum effect greatly.

\begin{figure}[thb]
\begin{center}
\includegraphics[width=2.2in]{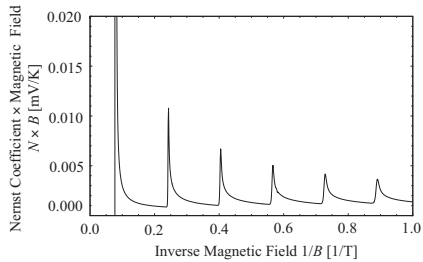}
\end{center}
\caption{The Nernst coefficient $N$ times the magnetic field, which is equal to 
the thermomagnetic power  $S_{xy}$, 
in three dimensions at $T = 0.28$ K against the inverse magnetic field $1/B$ [1/T], after Eq. (\protect\ref{nernstA}). 
%使用したパラメターのこと
We used the following parameter values:
the effective mass is 
$m= 0.06289 m_0$ and $m_z=0.6667 m_0$\protect\cite{effectiveMass}
 where $m_0$ is the bare mass of the electron; 
the size of the Hall bar is $L=4.0 {\rm mm} $\protect\cite{Behnia,BehniaScience} and 
$W=2.2 {\rm mm} $\protect\cite{Behnia,BehniaScience};
the confining potential at the edges is $V_0 = 4.22$ eV\protect\cite{edgeV}; 
the chemical potential at equilibrium is $\mu= 11.4 {\rm meV}$\protect\cite{chemicalP}.
}
\label{edge}
\end{figure}

%\begin{figure}[htb]
%%\begin{center}
%\includegraphics[width=2.2in]{nernstinv01.eps}
%%\end{center}
%\caption{ NOW PREPARING: 
%thermal conductance $G$ in three dimensions at $T=1[K]$ against the inverse magnetic field $1/B[1/T]$}
%\label{thermalcond}
%\end{figure}

%その他パラメーター設定
%me = 9.1093897*10^(-31);
%kB = 1.380658*10^(-23);
%qe = 1.60217733*10^(-19);
%\[HBar] = 1.05457266*10^(-34);
%m = 0.06289 me;
%mz = 0.6667 me;
%Ly = 2.2*10^(-3);
%dens = 9.75 * 10^3;
%vs = 2 * 10^3;
%\[Rho]xx = 5* 10^(-3);
%defp = 1.2 *qe;
%\[Mu] = 11.4*10^(-3)*qe;

%ネルンスト係数のピークが生き残ったこと　定性的に一致
%オーダーが２桁くらいずれていること
Figure \ref{edge} presents the quantum Nernst effect thus obtained in three dimensions.
This agrees with the experimental result qualitatively, though the magnitudes of the peaks is much smaller in our calculation than in the experimental result.

%On the other hand, in Fig.\ref{edge}(b)
%the plateau of thermal conductance, can be seen in two dimensional system, became unclear as Hall conductance in three dimensions.

%convect
%バンホーフ特異性のこと
%van Hove singularity
%状態密度がz方向に重なったこと
%These results are attributed to a singularity of the density of state.
%The density of states in three dimensions still shapes a singularity 
%because the density of state in x-y plain which shows the van Hove singularity is superimposed along $Z$ direction. 
%%示す
%%This fact is evidence of ~
%%indication
%Changing a space between two neighboring Landau levels  as the applied magnetic field increased,   
%when the chemical potential coincide with one of Landau levels, 
%the current is carried throughout the bulk.
%The peaks of Nernst coefficient in thermoelectrical current are at precisely those points.
%In contrast when it lies between a pair of neighbor Landau levels, 
%current is carried throughout edge states.

%%%%%フォノンドラッグ%%%%%%%%%%%%%%
\section{Phonon-drag effect on the Nernst coefficients}
%いまは短い論文なのでいらない
%In this section, we consider a phonon-drag effect to the Nernst coefficient, for agreement with the experiment quantitatively.
%A temperature gradient affects not only carriers but also phonons.
 
 %In the presence of temperature gradient not only carrier but also phonon  ？？？？
 %The thermomagnetic power  $S_g$
 
 Since the above argument does not quantitatively agree with the experimental result, we now consider the phonon-drag effect.
 We basically follow Sugihara's theory\cite{{Sugihara},{Sugihara2}} of the thermoelectric power in bismuth. %extending the calculation to the region of a strong magnetic field.
The difference here is that we treat the Fermi and Bose distribution functions without any approximations and evaluate the magnetic-field dependence of the thermomagnetic power  numerically.

The thermomagnetic power  $S_{xy} = NB$ is given by \begin{equation}
S_{xy}= - \frac{Q_y}{T F} \rho_{xx} ,
\end{equation}
where $Q_y$ denotes the heat current in the $y$ direction, $F$ denotes the electric field in the $x$ direction, $\rho_{xx}$ denotes the diagonal resistivity, and
 we assumed $\rho_{xy} \ll \rho_{xx}$ in bismuth.
At low temperatures  we may neglect all excitations except 
 acoustic phonons having the energy $\hbar \omega_q $ and the wave vector {\boldmath $q$}, which are generated through deformation coupling.
The heat current in the $y$ direction is then given by
\begin{equation}
Q_y = \int \frac{d \mbox{{\boldmath $q$}}}{(2\pi)^3}  \hbar \omega_q v_s \frac{q_y}{q} \left( N_q - N_q^{(0)}  \right) , 
\end{equation}
where $v_s$ is the group velocity of the phonons.
The term $N_q - N_q^{(0)}$ represents the displacement of the phonon distribution $N_q$ from its equiribrium Bose distribution $N_q^{(0)}$.

In order to estimate the displacement, we use the Boltzmann equation in the steady state
%%輸送方程式
\begin{equation}
\left( \frac{\partial N_q}{\partial t} \right)_{\rm carrier} + 
\left( \frac{\partial N_q}{\partial t} \right)_{\rm relaxation} = 0 .
\label{Boltzmann}
\end{equation}
The first term represents the change of the phonon distribution due to the interaction with carriers and the second term represents that due to other interactions such as boundary scattering, phonon-phonon interaction and impurity scattering.
In the steady state these two terms are balanced.
In the second term, 
we assume $(\partial N_q / \partial t)_{{\rm relaxation}} = - (N_q - N_q^{(0)})/\tau_{\rm r}(q)$ in the relaxation-time approximation; 
the electron-phonon interaction replaces the phonon distribution, but 
the relaxation effects make the nonequilibrium phonon distribution to the equilibrium one in time $\tau_{\rm r}$.

\begin{figure}[htb]
\begin{center}
\includegraphics[width=2.2in]{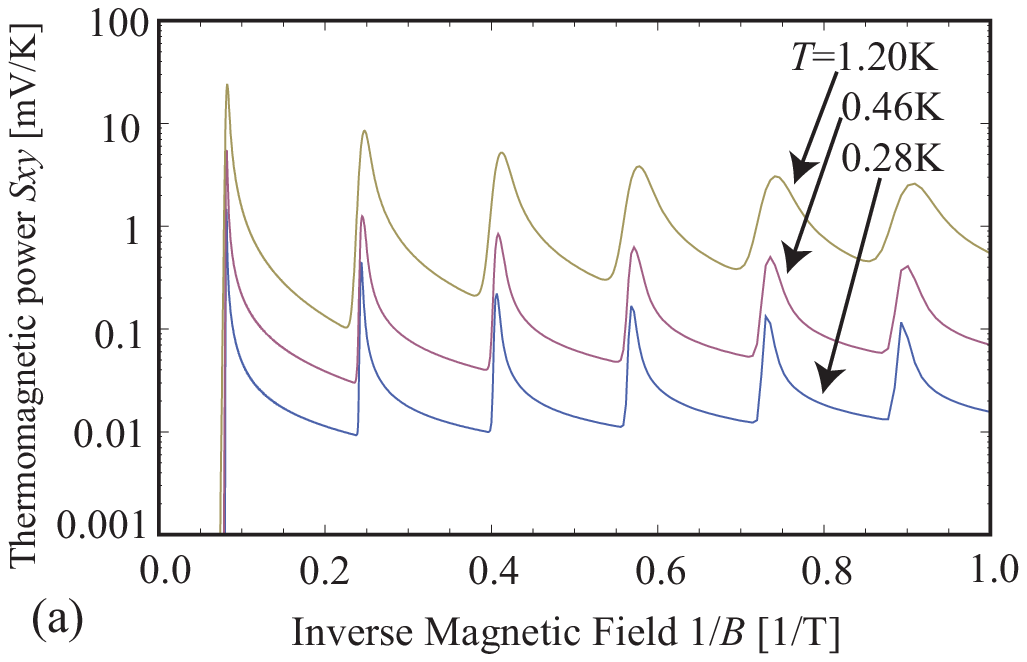}
\includegraphics[width=2.2in]{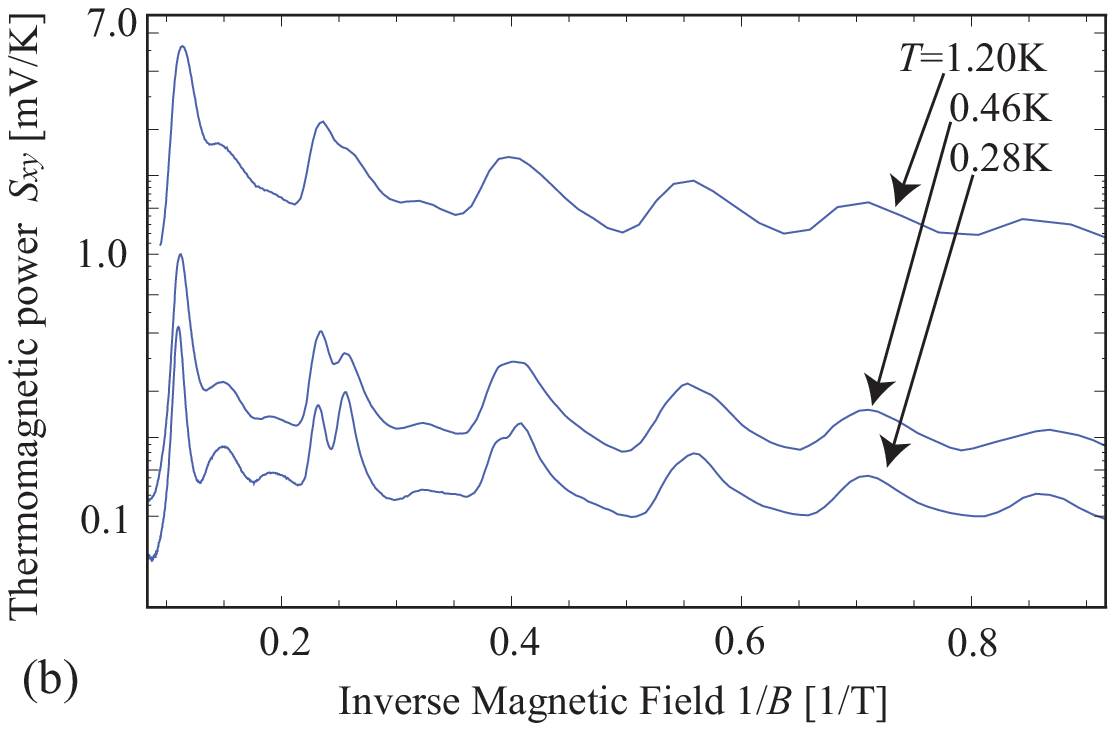}
\includegraphics[width=2.2in]{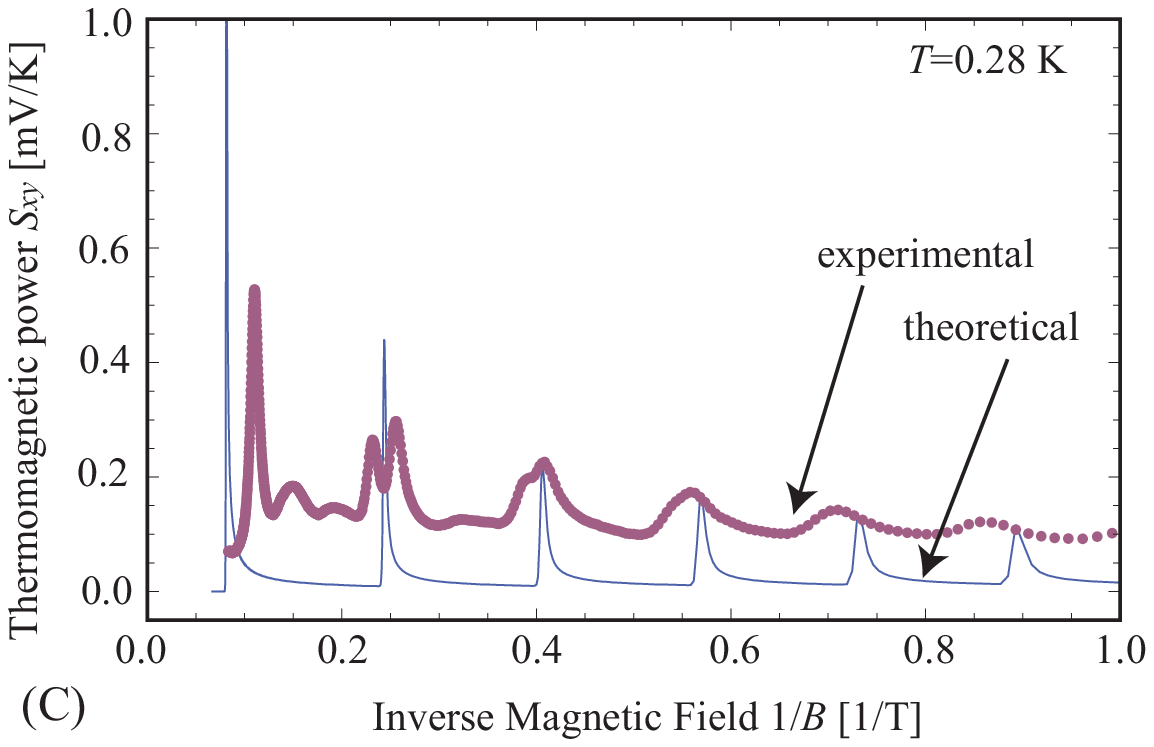}
\end{center}
\caption{The thermomagnetic power  in three dimensions at $T$= 0.28 K, 0.46 K, 1.20 K
 against the inverse magnetic field $1/B$[1/T]: 
 (a) Our theoretical result with the phonon-drag effect;  
 (b) The experimental result by Behnia {\it et al}.\protect\cite{Behnia,BehniaScience}; 
 (c) Comparison of the results at $T$= 0.28 K. 
 In the calculation we used the following parameter values: the deformation potential $D=1.2$eV\protect\cite{deform}; 
the electric resistivity $\rho_{xx}=0.35\Omega {\rm cm} $\protect\cite{Behnia,BehniaScience}; 
 the group velocity of the phonons $2 \times 10^3 {\rm m/s}$; the bismuth density $9.75\times 10^3 {\rm kg/m^3}$. The parameter values for the effective mass $m$ and $m_z$ and the chemical potential at equilibrium $\mu$ are the same as in Fig.~ 1.
}
\label{phonondrag}
\end{figure}

The theoretical result is shown in Fig.~\ref{phonondrag}~(a) and the experimental result is shown in Fig.~\ref{phonondrag} (b) at three temperatures, respectively.
 In Fig.~\ref{phonondrag} (c) we compared the theoretical and experimental result at T=0.28[K]. 
The magnitude of the peaks of the theoretical result, after considering the phonon-drag effect, is  consistent with the experimental result. 
Note that there are no adjustable fitting parameters.
%It can be seen that the ~.
%Peak magnitude of ~ along~ 
%while under the ~ condition these ~ did not significantly change.

%%%%%まとめ%%%%%%%%%%%%%%%%
\section{Summary}
We have shown that the quantum Nernst effect survive in three dimensions due to a van Hove singularity.
We have presented an extension of the edge-current picture in two dimensions to three dimensions, in order to explain Behnia {\it et al}.'s\cite{Behnia} experiment for a bismuth single crystal.
The naive extension reproduced the quantum oscillation of the experimental result qualitatively.
%We have found that the peaks are smaller than the experimental values of thermomagnetic power  by 2 order of magnitude. こういう否定的な言い方をすると何を言いたいのか理解されない
Furthermore, consideration of the phonon-drag effect led us to a theoretical result quantitatively consistent with the experimental result.
% we found the peaks of the Nernst coefficients survive in three dimensions though the order  is less than Behnia et al.'s experiment.
%And then, we consider the difference between theory and experiment is due to some interaction affects thermal transport, and assume the major interaction at low temperatures is carrier-phonon interaction\cite{Farag}. 
The peaks are sharper than those in the experiment 
probably because we neglected impurity scattering of carriers.  
For simplicity, we also assumed the Fermi surface of bismuth is ellipsoid and 
considered only the hole contribution.
Going beyond these approximations would be an interesting future problem.

%量子ホール効果でも、端電流描像と、計算するのが一致する。それに対応している。（？？？）

%他の研究との比較\\
%Finally, we  comment on another recent approach to the thermal transports.
%T.Yamamoto et al. studied thermal conductances quantized due to phonons in carbon nanotubes.
%We considered thermal transport through interactions between carrier and phonon. 
%おしまい？

 %%%%%%%%%謝辞%%%%%%%%%%%%%
 \section*{Acknowledgments}
We thank Dr Y. Hasegawa for constructive comments.
We also thank Dr. Behnia for valuable comments and providing experimental data.
The work is supported partly by the Murata Science Foundation as well as by
the National Institutes of Natural Sciences undertaking Forming Bases for Interdisciplinary and International Research through Cooperation Across Fields of Study and Collaborative Research Program (No.~NIFS08KEIN0091) and Grants-in-Aid for Scientific Research (No.~17340115 and No.~20340101) from the Ministry of Education, Culture, Sports, Science and Technology.

%%%%かきかた%%%%%%%%%%%%%%%%%%%%%%
\balance

%%%%%%%%%%%%%%%%%%%%%%%%%%
%%%%参考文献%%%%%%%%%%%%%%%%%

\end{document}